\documentclass[11pt]{article}
\usepackage{amsmath}
\usepackage{diagram}
\usepackage{tikz}
\usetikzlibrary{arrows}
\usepackage{graphicx}
\usepackage{amsfonts}
\usepackage{amssymb}
\usepackage{multirow}
\def\Proof.{{\bf{Proof. }}}

\begin{document}

\title{ An upper bound for the number of chess diagrams \\  without promotion }
\author{Daniel GOURION\footnote{  Laboratoire de Mathematiques d'Avignon, Avignon Universite, France.}}
\date{October 2022}
%Michel VOLLE\footnote{  University of Avignon, 33, rue
%Louis Pasteur, 84000 Avignon, France.} }
\maketitle

\vspace{-0.5cm}

\begin{abstract}
In 2015, Steinerberger showed that the number of legal chess diagrams without promotion is bounded from above by $2\times 10^{40}$. This number was obtained by restricting both bishops and pawns position and by a precise bound when no chessman has been captured. We improve this estimate and show that the number of legal diagrams is less than $4\times 10^{37}$. To achieve this, we define a graph on the set of diagrams and a notion of class of pawn arrangements, leading to a method for bounding pawn positions with any number of men on the board. 
\end{abstract}

\vspace{0.5cm}
\noindent\textbf{ACKNOWLEDGEMENT. This manuscript has been accepted for publication in ICGA Journal, DOI: 10.3233/ICG-220210} 

%\begin{keyword}
%\kwd{Shannon's number}
%\kwd{Chess}
%\kwd{State space}
%\end{keyword}

%%%%%%%%%%% The article body starts:

\section{Introduction}\label{s1}

The state space of chess is the set of all possible configurations of a chess game. It gives an estimate of the computational complexity of the game. Unfortunately, chess configurations are less easy to define than chess games. Following the definitions given by Labelle (2011), we will call a diagram the contents of the 64 squares of the chessboard. There are six types of chessmen: pawn, knight, bishop, rook, queen and king. Each can be white or black, thus a square may contain one of 12 possible chessmen or  be empty resulting in 13 possibilities for the content of each square. 

Taking also into account whose turn it is, castling rights, and any en passant square, we define what we call a position. Thus a position is a diagram with three kinds of additional information. The first one is whether it is white or black to move. The second one is whether it is possible for each side to castle kingside or queenside. The last one states any possible en passant target square. A position can be easily encoded using for example ``Forsyth–Edwards Notation'' (2003), see Fig. \ref{f1}b for an example.

A diagram (resp. a position) is legal if it is reachable from the starting position in a valid game of chess. For example, diagrams or positions with either pawns on ranks 1 and 8, more than 32 pieces on the board, both kings in check or two neighbouring kings are illegal. The same is true for positions with one king in check and the other side to move.  There are a lot of other illegal diagrams and positions and it is impossible to describe all of them. Indeed, proving illegality of a chess diagram or position is most of the time not an easy task. Figure \ref{f1} shows three examples: a legal diagram, a legal position, an illegal diagram. Note that all positions encompassing an illegal diagram are illegal.

In the sequel, the expression pawn (resp. piece) arrangement will be used to refer to the way to place only pawns (resp.  pieces) on the chessboard, whether empty or not. A pawn (resp. piece) arrangement is illegal if no legal diagram exists with this pawn (resp. piece) arrangement. For example, Fig. 1c shows a position whose  pawn arrangement is illegal. Note that the term pieces here is excluding pawns. We will use the term men or chessmen for the physical pieces of the set, including the pawns.

\begin{figure}
\begin{diagram}
\setboolean{piececounter}{false}
\setboolean{showcomputer}{false}
\specialdiagnum{a.}
\fen{r3k2r/p1pqbpp1/2npbn2/Pp2p3/2BPP1Pp/1PN1BN2/2PQ1P1P/R3K2R}
\end{diagram}
\begin{diagram}
\setboolean{piececounter}{false}
\setboolean{showcomputer}{false}
\specialdiagnum{b.\quad     w KQq b6}
\fen{r3k2r/p1pqbpp1/2npbn2/Pp2p3/2BPP1Pp/1PN1BN2/2PQ1P1P/R3K2R}
\end{diagram}
\begin{diagram}
\setboolean{piececounter}{false}
\setboolean{showcomputer}{false}
\specialdiagnum{c.}
\fen{r1bqkbnr/pppp1ppp/8/8/3NP3/3P3P/PPP3PP/RNBQKB1R}
\end{diagram}

\caption{Left. A legal diagram. Middle. A legal position: white is to move, white may castle kingside or queenside, black may castle queenside, axb6 en passant is allowed. This information is indicated respectively by w, K, Q, q and b6, using Forsyth–Edwards Notation. Right. An illegal diagram: where does the h3 pawn come from?}
\label{f1}
\end{figure}

The paper is structured as follows. In Section 2 the existing literature on the number of legal diagrams and positions is summarized. In Section 3 is described a method to compute an upper bound for the number of legal diagrams. The case with 24 men or more is handled from subsections \ref{s21} to \ref{s25}. Several tools are combined: we define a partition on the set of diagrams, create a graph on this partition, and define a notion of class of pawn arrangements. Then is described the possible consequences of any capture on a pawn arrangement. Starting from diagrams with 32 chessmen and decreasing progressively the number of chessmen on the board, we calculate at each step an upper bound on the number of legal pawn and piece arrangements. In subsection \ref{s26} is treated the case with 23 men or less. In Section 4, detailed results are shown and in Section 5 possible improvements and generalizations are proposed.

\section{Related Research}

In a famous paper Shannon (1950) estimated the number of legal diagrams of the order of $64!/32!(8!)^2(2!)^6$ $\approx4.63\times 10^{42}$. This is, of course, a very rough estimate: it does not consider that some legal diagrams have less than 32 men on the board. Neither does it take into account the fact that some pawns could be promoted. These two factors lead to an underestimation of the number of diagrams. On the other hand, this estimate accounts for a lot of illegal diagrams, most importantly illegal pawn arrangements.

Chinchalkar (1996) proved an upper bound on the number of legal positions of approximately $1.78\times 10^{46}$. More recently, Tromp (2021b) claims an upper bound of $8.73\times 10^{45}$. He then used 1 000 000 randomly generated positions and checked their legality, yielding an estimated number of legal positions of $(4.79 \pm 0.04)  \times 10^{44}$ with 95\% confidence level.

As for the diagrams, Steinerberger (2015) improved Shannon's number, giving an upper bound of approximately $1.53\times 10^{40}$ for the number of legal chess diagrams without promotion. For doing so, he used the fact that some men can not occupy any square of the board: each bishop is either light-squared or dark-squared, and pawns can not be located on rank 1 or 8. Moreover, he treated separately the case with 32 men on the board, reducing further the number of pawn arrangements in that case. Tromp (2021b) improved this estimate down to approximately $2.89\times 10^{39}$ by constraining the number of unopposed pawns according to pawns and pieces captured.

In the present work, we drastically reduce the combinatorial complexity of the setup of pawns when 24 to 32 men are on the board. For the cases with less than 24 men on the board, we propose a way to arrange bishops and kings before adding the pawns and the remaining men. These efforts result in an upper bound of less than $4\times 10^{37}$ for the number of legal diagrams without promotion.

\section{Methods}\label{s2}

\subsection{A graph on the set of diagrams}\label{s21}
Each diagram is associated with a unique quadruplet $\mathcal{P}=(P_w,P_b,p_w,p_b)$ where $P_w$ and $P_b$ are respectively the number of white and black pieces and $p_w$ and $p_b$ are the number of white and black pawns. As we are interested in the case without promotion, the following inequalities hold: $1\leq P_w\leq 8$, $1\leq P_b\leq 8$, $0\leq p_w\leq 8$ and $0\leq p_b\leq 8$. A given quadruplet represents a subset of the set of diagrams and the family of all quadruplets is a partition of the set of diagrams.

The set of quadruplets can be represented as an oriented graph: each quadruplet is a vertex of the graph and edges are the legal transitions during a game of chess from a quadruplet to another (see Fig.\ref{f2}). These transitions are only due to captures, as we do not take promotions into account. For each edge of the graph, the color and the nature (piece or pawn) of the captured piece is known. Hence we also know the color of the capturing man, but most of the time its nature is unknown, as it can be either a piece or a pawn. On the graph, any quadruplet except the root has at least one and at most four predecessors. For example $(8,8,8,6)$ has just one predecessor which is $(8,8,8,7)$, whereas $(7,6,7,5)$ has four predecessors which are $(8,6,7,5)$, $(7,7,7,5)$, $(7,6,8,5)$ and $(7,6,7,6)$. The edge between $(8,8,8,7)$ and $(8,8,8,6)$ represents the capture of a black pawn. We define the root of the graph as the quadruplet $\mathcal{P}_0=(8,8,8,8)$. The diagram of the starting position, as well as all the diagrams with 32 men, is associated with this quadruplet.

\begin{figure}[h!]

\begin{tikzpicture}

\tikzset{vertex/.style = {shape=circle,draw,minimum size=1.5em}}
\tikzset{edge/.style = {->,> = latex'}}

\node[vertex] (a) at  (7,3) {(8,8,8,8)};
\node[vertex] (b1) at  (1,1) {(8,8,8,7)};
\node[vertex] (b2) at  (5,1) {(8,8,7,8)};
\node[vertex] (b3) at  (9,1) {(8,7,8,8)};
\node[vertex] (b4) at  (13,1) {(7,8,8,8)};
\node[vertex] (c1) at  (0,-3) {\footnotesize(8,8,8,6)};
\node[vertex] (c2) at  (1.5,-3) {\footnotesize(8,8,7,7)};
\node[vertex] (c3) at  (3,-3) {\footnotesize(8,7,8,7)};
\node[vertex] (c4) at  (4.5,-3) {\footnotesize(7,8,8,7)};
\node[vertex] (c5) at  (6,-3) {\footnotesize(8,8,6,8)};
\node[vertex] (c6) at  (7.5,-3) {\footnotesize(8,7,7,8)};
\node[vertex] (c7) at  (9,-3) {\footnotesize(7,8,7,8)};
\node[vertex] (c8) at  (10.5,-3) {\footnotesize(8,6,8,8)};
\node[vertex] (c9) at  (12,-3) {\footnotesize(7,7,8,8)};
\node[vertex] (c10) at  (13.5,-3) {\footnotesize(6,8,8,8)};
%edges
\draw[edge] (a) to (b1);
\draw[edge] (a) to (b2);
\draw[edge] (a) to (b3);
\draw[edge] (a) to (b4);
\draw[edge] (b1) to (c1);
\draw[edge] (b1) to (c2);
\draw[edge] (b1) to (c3);
\draw[edge] (b1) to (c4);
\draw[edge] (b2) to (c2);
\draw[edge] (b2) to (c5);
\draw[edge] (b2) to (c6);
\draw[edge] (b2) to (c7);
\draw[edge] (b3) to (c9);
\draw[edge] (b3) to (c8);
\draw[edge] (b3) to (c6);
\draw[edge] (b3) to (c3);
\draw[edge] (b4) to (c10);
\draw[edge] (b4) to (c9);
\draw[edge] (b4) to (c7);
\draw[edge] (b4) to (c4);
\end{tikzpicture}
\caption{Subgraph for 32, 31 and 30 chessmen.}\label{f2}
\end{figure}

\subsection{Classes of pawn arrangements}\label{s22}

The most important improvement of our method is to compute a better upper bound for the number of legal pawn arrangements. For that purpose, we define a class of pawn arrangements as follows: a class is an array of 8 columns. Each column represents a file of the chessboard (first column is the a file,...and last column is the h file) and contains between 0 and 6 elements. The length of a column is the number of pawns on the file and the elements of this column are either $w$ for white pawns or $b$ for black pawns. The order of elements in a column is important: they are written according to their distance to the 7th rank of the chessboard. Thus the element of the first row of a given column is the closest pawn to the 7th rank and the element of the last row is the closest pawn to the second rank. Note that because no pawn can occupy the first and last rank of the board, a column can not contain more than 6 elements.

For each class of pawn arrangements, it is easy to determine an upper bound for the number of possible arrangements of black and white pawns. Let us denote by $k_i$, $i=1,\ldots,8$ the number of elements of column $i$. These $k_i$ pawns can be placed on 6 squares. Since we know the relative position of white and black pawns on each file, the number of pawn arrangements for a given class is equal to $\prod_{i=1}^{8}\binom{6} {k_i}$. Knowing the class of pawn arrangements of a diagram does not indicate the square occupied by a given pawn, but only its position relatively to pawns on the same file. Hence, a great number of different pawn arrangements are encoded in the same class. This makes possible the computation and storage of all the legal classes of a given quadruplet, while the same would be impossible for legal pawn arrangements.

The class of pawn arrangements is modified during a game of chess only in case of a capture or a promotion. Other pawn moves do not change it.  For this reason, the class of pawn arrangements of any legal diagram associated with $\mathcal{P}_0=(8,8,8,8)$ is the same as the class of the starting position, given in Table \ref{t1} (the first row of each column is the upper in the table and contains the closest pawn to the 7th rank, the last row is the lower and contains the closest pawn to the second rank). Hence, an upper bound for the number of different arrangements for white and black pawns for this class is $\prod_{i=1}^{8}\binom{6} {2}=15^8\approx 2.56e+9$.

\begin{table*}[h!]
\caption{The unique class of pawn arrangements for 32 men legal diagrams} \label{t1}
\begin{center}
\begin{tabular}{c c c c c c c c} 
 \hline
 Col 1 & Col 2 & Col 3 & Col 4 & Col 5 & Col 6 & Col 7 & Col 8 \\ [0.5ex] 
 \hline
 b & b & b & b & b & b & b & b\\ 

 w & w & w & w & w & w & w & w\\ 
 \hline
\end{tabular}
\end{center}
\end{table*}

We illustrate the notions of quadruplets and classes of pawn arrangements by the first moves of a chess game (see Fig. \ref{f3} and Tables \ref{t1} and \ref{t2}). In the first diagram, the 32 chessmen are still on the board, thus this diagram is represented by the quadruplet $(8,8,8,8)$ and the class of pawn arrangements of Table \ref{t1}. In the second diagram the white d-pawn has been captured by the black e-pawn, leading to the quadruplet $(8,8,7,8)$ and the class of pawn arrangements of Table \ref{t2}, upper part. In the next diagram a white knight has captured a black pawn on d4 leading to the quadruplet $(8,8,7,7)$ and the class of pawn arrangements of Table \ref{t2}, middle part. The next capture is when a white knight take a black knight, after which the quadruplet of the arising diagram is $(8,7,7,7)$ and the class of pawn arrangements is unchanged. Then a black pawn on b7 recaptures the white knight on c6, leading to the quadruplet $(7,7,7,7)$ and the class of pawn arrangements of Table \ref{t2}, lower part.

\begin{figure}[h!]
\begin{diagram}
\setboolean{piececounter}{false}
\setboolean{showcomputer}{false}
\specialdiagnum{a.}
\fen{r1bqkbnr/pppp1ppp/2n5/4p3/3PP3/5N2/PPP2PPP/RNBQKB1R}
\end{diagram}
\begin{diagram}
\setboolean{piececounter}{false}
\setboolean{showcomputer}{false}
\specialdiagnum{b.}
\fen{r1bqkbnr/pppp1ppp/2n5/8/3pP3/5N2/PPP2PPP/RNBQKB1R}
\end{diagram}
\begin{diagram}
\setboolean{piececounter}{false}
\setboolean{showcomputer}{false}
\specialdiagnum{c.}
\fen{r1bqkbnr/pppp1ppp/2n5/8/3NP3/8/PPP2PPP/RNBQKB1R}
\end{diagram}
\begin{diagram}
\setboolean{piececounter}{false}
\setboolean{showcomputer}{false}
\specialdiagnum{d.}
\fen{r1bqkb1r/pppp1ppp/2N2n2/8/4P3/8/PPP2PPP/RNBQKB1R}
\end{diagram}
\begin{diagram}
\setboolean{piececounter}{false}
\setboolean{showcomputer}{false}
\specialdiagnum{e.}
\fen{r1bqkb1r/p1pp1ppp/2p2n2/8/4P3/8/PPP2PPP/RNBQKB1R}
\end{diagram}

\caption{Five legal diagrams, in a game beginning with the moves: 1. e4 e5 2. Nf3 Nc6 3. d4 (top left) exd4 (middle top) 4. Nxd4 (top right) Nc6 5. Nxc6 (bottom left) bxc6 (bottom right). }
\label{f3}
\end{figure}

\begin{table*}[h!]
\caption{Three classes of pawn arrangements relative to the second, third and last diagrams of Fig 3.} 
\label{t2}
\begin{center}
\begin{tabular}{c c c c c c c c} 
 \hline
 Col 1 & Col 2 & Col 3 & Col 4 & Col 5 & Col 6 & Col 7 & Col 8 \\ [0.5ex] 
 \hline
 b & b & b & b & w & b & b & b\\ 

 w & w & w & b &  & w & w & w\\ 
 \hline
\end{tabular}
\vspace{0.2cm}

\begin{tabular}{c c c c c c c c} 
 \hline
 Col 1 & Col 2 & Col 3 & Col 4 & Col 5 & Col 6 & Col 7 & Col 8 \\ [0.5ex] 
 \hline
 b & b & b & b & w & b & b & b\\ 

 w & w & w &  &  & w & w & w\\ 
 
 \hline
\end{tabular}
\vspace{0.2cm}

\begin{tabular}{c c c c c c c c} 
 \hline
 Col 1 & Col 2 & Col 3 & Col 4 & Col 5 & Col 6 & Col 7 & Col 8 \\ [0.5ex] 
 \hline
 b & w & b & b & w & b & b & b\\ 

 w &  & b &  &  & w & w & w\\ 
 
   &  & w &  &  &  &  & \\ 
 
 \hline
\end{tabular}
\end{center}
\end{table*}

\newpage

\subsection{Effect of a capture on a class of pawn arrangements}\label{s23}

In order to compute the classes of pawn arrangements of a given quadruplet, we describe the effect of a capture on a class. Without loss of generality, suppose that the captured man is white. There are four kinds of capture. Here are the consequences of each one:
\begin{itemize}
\item when a black piece takes a white piece, the class of pawn arrangements is unchanged.
\item when a black pawn takes a white piece: remove $b$ in the appropriate column of the array and insert $b$ in the appropriate adjacent column. 
\item when a black piece takes a white pawn: remove $w$ in the appropriate column of the array.
\item when a black pawn takes a white pawn: for the appropriate couple $(b,w)$ in adjacent columns, remove $b$ from its column and replace $w$ with $b$.  
\end{itemize}

For instance, take a legal diagram with 32 men. Its class of pawn arrangements is given in Table \ref{t1}. Observe the possible classes resulting from the capture of a white pawn.  We obtain 8 classes if the pawn is captured by a black piece (one for each possibility of removing $w$ in the array) and 14 classes if the pawn is captured by a black pawn (one for each couple $(b,w)$ in adjacent columns of the array). Now observe what happens with the same initial diagram when a white piece is captured. One class is generated when the capture is made by a black piece. If a black pawn captures a white piece, there are 14 possibilities for choosing adjacent columns and for each one, 3 different places where inserting $b$, generating $14\times 2$ different classes after removing duplicates.

\subsection{An upper bound for the number of legal pawn arrangements of a quadruplet}

 Without promotion, only captures can change the class of pawn arrangements during a game. Consequently, the list of classes of a given quadruplet $\mathcal{P}$ is inherited from the list of classes of its predecessors in the graph. Moreover, the list of class of the root $\mathcal{P}_0=(8,8,8,8)$ is known. Note that it contains a unique class. Hence the list of classes of any quadruplet may be computed in a recursive way, starting from the root.

For any quadruplet $\mathcal{P}$ with 31 to 24 chessmen the following program is executed: 
\begin{itemize}
    \item Compute the predecessors of $\mathcal{P}$.
    \item For each class of pawn arrangements of each predecessor $\mathcal{P'}$.
    \begin{itemize}
        \item Compute and store all the possible classes of pawn  arrangements generated by the captures from $\mathcal{P'}$ to $\mathcal{P}$.
    \end{itemize}
    \item Remove duplicates from the list of classes.
    \item Compute the number of pawn arrangements of each class of $\mathcal{P}$.
    \item Compute the number of pawn arrangements $n_\mathcal{P}$ of $\mathcal{P}$.
\end{itemize} 

For example (see subsection \ref{s23}), this algorithm generates 22 classes for $\mathcal{P}=(8,8,7,8)$, resulting in an upper bound of $$n_\mathcal{P}=8\times \binom{6}{1} \times \binom{6}{2}^7+ 14\times\binom{6}{1} \times \binom{6}{2}^7\approx 2.26\times 10^{10},$$ and 29 classes for $\mathcal{P}=(7,8,8,8)$ with an upper bound of $$n_\mathcal{P}=  \binom{6}{2}^8+ 28\times\binom{6}{1} \times \binom{6}{3} \times \binom{6}{2}^6\approx 4.08\times 10^{10}.$$
Note that the method takes all possible cases of capture into account for upper bounding the number of legal pawn arrangements, but that some of the generated pawn arrangements may be illegal, see Section \ref{s4} for details.

Symmetry was used in order to reduce the number of computations: the number of pawn arrangements of $(P_w,P_b,p_w,p_b)$ and $(P_b,P_w,p_b,p_w)$  are equal. The computation time and the number of classes of a quadruplet are quickly increasing from 31 to 24 chessmen. For example $(8,8,8,7)$ has 22 classes, whereas $(5,5,7,7)$ has 31265498. Computing the classes of the former takes less than a second and the latter more than eight hours. Four quadruplets represent all the legal diagrams with 31 men, whereas 84 quadruplets were needed to compute the whole 24 men case. Moreover, the improvement generated by this method of counting pawn arrangements over the combinatorial method in \cite{r5} is decreasing at the same time: for 32 chessmen, the upper bound on the number of legal diagrams is divided by $10^7$, whereas this ratio is approximately $1.7$ for 24 men. For these reasons this way of counting pawn arrangements was not used for diagrams with less than 24 chessmen.

\subsection{Counting pieces arrangements}\label{s25}

Once the upper bound on the number of pawn arrangements $n_\mathcal{P}$ for a given quadruplet $\mathcal{P}=(P_w,P_b,p_w,p_b)$ is known, pieces are placed on the board. Let $n=64-p_w-p_b$ be the number of unoccupied squares. When placing bishops, we take advantage of the fact that two bishops of the same side are placed on squares of different color. Hence the number of arrangements for these two bishops is bounded by $n^2/4$ when $n$ is even and $(n^2-1)/4$ when $n$ is odd. Let $b_w$ and $b_b$ be respectively the number of white and black bishops.

The following algorithm is applied:

\begin{itemize}
    \item Find all combinations containing $P_w$ white pieces and $P_b$ black pieces
    \item For each combination
    \begin{itemize}
        \item Compute the number of possible arrangements of bishop pair(s), if any, on the $64-p_w-p_b$ squares.
        \item Compute the number of possible arrangements of remaining bishops, if any.
        \item Compute the number of possible arrangements of remaining pieces on the $64-p_w-p_b-b_w-b_b$ squares.
        \item Multiply these numbers to compute the number of piece arrangements of the combination.
    \end{itemize}
    \item Add these numbers to compute the number of piece arrangements $m_\mathcal{P}$ of $\mathcal{P}$
\end{itemize}

The upper bound on the number of legal diagrams of $\mathcal{P}$ is equal to $n_\mathcal{P}\times m_\mathcal{P}$. 

Please note that we did not exclude neighbouring kings in this count. There is no easy way to do so, the reason being that pawns have been placed before the kings.

\subsection{Diagrams with less than 24 elements}\label{s26}

For diagrams with 2 to 23 chessmen the same method as in \cite{r5} is used, with a slight refinement consisting in not letting kings occupy adjacent squares. First the kings and the bishops are placed on the board, considering how many of the 16 squares of the first and last rank (denote $A$ the set of these squares) they occupy and using the fact that kings may not be adjacent to each other. For any $0\leq i \leq 6$, let $f_i(b_w,b_b)$ denote the number of ways to place the kings, $b_w$ white bishops and $b_b$ black bishops on the board such that $i$ of these pieces are contained in $A$. The results are stored in Table \ref{t3}. Every case has been calculated by combinatorial arguments and checked with computer enumeration of all possibilities.

\begin{table*}[h]
\caption{Counting kings and bishops in first and last rank} \label{t3}
\begin{tabular}{c c c c } 
 \hline
 $f_0(.,.)$ & 0 & 1 & 2  \\ [0.5ex] 
 \hline
 0 & 1952 & 89792 & 1031644 \\ 

 1 & 89792 & 4040640 & 45392336 \\ 
 
 2 & 1031644 & 45392336 & 498806704 \\
 \hline
\end{tabular}
\qquad
\begin{tabular}{c c c c } 
 \hline
 $f_1(.,.)$ & 0 & 1 & 2  \\ [0.5ex] 
 \hline
 0 & 1448 & 99288 & 1517632 \\ 

 1 & 99288 & 6003920 & 84799744 \\ 
 
 2 & 1517632 & 84799744 & 1130721152 \\
 \hline
\end{tabular}

\smallskip

\begin{tabular}{c c c c } 
 \hline
 $f_2(.,.)$ & 0 & 1 & 2  \\ [0.5ex] 
 \hline
 0 & 212 & 31896 & 757472 \\ 

 1 & 31896 & 2988432 & 57608192 \\ 
 
 2 & 757472 & 57608192 & 978967872 \\
 \hline
\end{tabular}
\qquad
\begin{tabular}{c c c c } 
 \hline
 $f_3(.,.)$ & 0 & 1 & 2  \\ [0.5ex] 
 \hline
 0 & 0 & 2968 & 152320 \\ 

 1 & 2968 & 589008 & 17763648 \\ 
 
 2 & 152320 & 17763648 & 413211008 \\
 \hline
\end{tabular}

\smallskip

\begin{tabular}{c c c c } 
 \hline
 $f_4(.,.)$ & 0 & 1 & 2  \\ [0.5ex] 
 \hline
 0 & 0 & 0 & 10276 \\ 

 1 & 0 & 38584 & 2473408 \\ 
 
 2 & 10276 & 2473408 & 89297152 \\
 \hline
 \end{tabular}
 \qquad
\begin{tabular}{c c c c } 
 \hline
 $f_5(.,.)$ & 0 & 1 & 2  \\ [0.5ex] 
 \hline
 0 & 0 & 0 & 0 \\ 

 1 & 0 & 0 & 123312 \\ 
 
 2 & 0 & 123312 & 9324672 \\
 \hline
 \end{tabular}
 \qquad
 
 \smallskip
 
 \begin{tabular}{c c c c } 
 \hline
 $f_6(.,.)$ & 0 & 1 & 2  \\ [0.5ex] 
 \hline
 0 & 0 & 0 & 0 \\ 

 1 & 0 & 0 & 0 \\ 
 
 2 & 0 & 0 & 364560 \\
 \hline
\end{tabular}
\end{table*}

Some examples of such calculation are given. We begin with only kings on the board. If the white king is in a corner, the black king can occupy 14 squares in $A$. If the white king is in $A$ but not in a corner, the king can occupy 13 squares in $A$. For this reason $f_2(0,0)=4\times 14 + 12\times 13 = 212$. We obtain in the same way $f_0(0,0)$ and $f_1(0,0)$ giving the number of legal diagrams with only kings on the board with respectively 0 king and 1 king in $A$.

Now we describe a more complicated scenario. The proof of the other cases is left to the reader. $f_1(2,2)$ requires 4 bishops on the board and 1 piece in $A$. Hence there are two cases. In the first case, one king is in $A$, the other one and the 4 bishops are not in $A$. As seen above, we first place both kings. There are $f_1(0,0)=1448$ ways to place them. Consider the color of the square occupied by the king outside $A$. Two bishops are on squares of this color, yielding $23\times 22$ ways to arrange them, and two bishops are on different color, producing $24\times 23$ cases. Hence,
this first case yields $1448\times 24 \times 23^2 \times 22$ possibilities. In the second case, no king is in $A$. We infer that we have a bishop in $A$ and other pieces outside $A$. First we place the kings, there are $f_0(0,0)=1952$ ways to place them outside $A$. We consider two subcases. In the first one, the kings are on opposite colors, that is 988 cases, in the second one the kings are on the same color, that is 964 cases. When the kings are on opposite colors, first is chosen which bishop is inside $A$ and on which square it is, yielding $4\times 8$ cases. Next we arrange the bishops outside $A$, obtaining $23^2 \times 22$ ways of placing them. When the kings are on the same color, we have to distinguish whether the bishop in $A$ is on the same color as the kings or not. If it is, with the same method we obtain $2\times 8\times 24\times 23\times 22$ ways of placing the four bishops. If the bishop inside $A$ is not on the same color as the kings, we obtain $2\times 8\times 24\times 22\times 21$ ways of placing the four bishops. Therefore

\begin{multline*}
f_1(2,2)=1448\times 24 \times 23^2 \times 22+988\times4\times 8\times 23^2 \times 22\\ +964(2\times 8\times24 \times 23\times 22+ 2\times 8\times24 \times 22\times 21)=1130721152.
\end{multline*}

Once Table \ref{t3} is computed, we proceed by a rather large case distinction. Every case has a simple combinatorial structure. For the sake of completeness, we sum up the algorithm proposed in \cite{r5}, slightly modified as described:

\begin{itemize}
    \item For all possible values of $k$ kings and bishops contained in $A$, $0 \leq  k \leq 6$, 
    \begin{itemize}
        \item For all possibles values of $(b_w,b_b)$, there are $f_k(b_w,b_b)$ ways of placing two kings, $b_w$ and $b_b$ bishops on the board with precisely $k$ of them in $A$.
        \begin{itemize}
            \item There are $48-(2+b_w+b_k-k)$ squares on which the $p_w$ white pawns are placed and $48-(2+b_w+b_b-k)-p_w$ on which the black pawns are placed.
            \begin{itemize}
                \item There are $62-b_w-b_k-p_w-p_k$ squares on which to place the remaining men. 
            \end{itemize}
        \end{itemize}
    \end{itemize}
\end{itemize}
The upper bound for each case is simply a product of $f_k(b_w,b_b)$ and binomial coefficients. Then all these numbers are summed over all cases containing the same number of chessmen to generate an upper bound for the number of legal diagrams with 2 to 23 men on the board. This method yields also a bound when there are 24 men or more on the board but it is far less accurate than the one found by the method based on pawn arrangements and quadruplets.

\section{Results}\label{s3}

An upper bound has been obtained for the number of legal diagrams of all quadruplets containing 24 to 32 chessmen (subsection \ref{s25}) and for diagrams with 2 to 23 chessmen  (subsection \ref{s26}). Summing over all possibilities yields that the upper bound for the total number of legal chess diagrams without promotion is equal to
$$
3.7422\ldots\times 10^{37}.
$$
Some details can be found in Table \ref{t4} for legal diagrams with 23 men or more. We observe that the upper bound for the number of diagrams reaches a maximum when there are 26 or 27 men on the board. The improvement factor, defined as the ratio between the upper bounds of the method described in \cite{r5} and our method is increasing with the number of chessmen, except for 32 chessmen. The improvement factor is particularly important in the 31 men case for which it exceeds 400 000. For 32 chessmen, it is only 3.83 because in this case both methods are equivalent for counting pawn arrangements.

\begin{table*}[h]
\caption{Upper bound on the number of legal diagrams for 23 to 32 men} \label{t4}
\begin{center}
\begin{tabular}{c c c c c c} 
 \hline
 & 32 men & 31 men & 30 men & 29 men & 28 men   \\ [0.5ex] 
 \hline
This method & $4.93\times 10^{32}$ & $1.71\times 10^{34}$ & $1.64\times 10^{35}$ & $1.53\times 10^{36}$ & $5.46\times 10^{36}$ \\ 

Steinerberger paper & $1.89\times 10^{33}$ & $6.97\times 10^{39}$ & $4.73\times 10^{39}$ & $2.29\times 10^{39}$ & $8.75\times 10^{38}$ \\

Improvement factor & 3.83 & 408000 & 28800 & 1490 & 160 \\ 
 \hline
 
\end{tabular}

\smallskip
\begin{tabular}{c c c c c c} 
 \hline
 &  27 men & 26 men & 25 men & 24 men & 23 men \\ [0.5ex] 
 \hline
This method &  $1.05\times 10^{37}$ & $1.08\times 10^{37}$ & $6.14\times 10^{36}$ & $2.09\times 10^{36}$ & $5.66\times 10^{35}$\\ 

Steinerberger paper & $2.78\times 10^{38}$ & $7.50\times 10^{37}$ & $1.75 \times 10^{37}$ & $3.54\times 10^{36}$ & $6.29\times 10^{35}$ \\

Improvement factor & 26.4 & 6.92 & 2.85 & 1.70 & 1.11\\ 
 \hline
 
\end{tabular}
\end{center}
\end{table*}

Computation time is less than 1 second for each quadruplet containing 30 men or more. For 28 men, it is less than 1 second for $(8,8,8,4)$, which has 2682 classes of pawn arrangements ($n_\mathcal{P}\approx 1.28 \times 10^{11}$) and 9 seconds for $(7,7,7,7)$, which has 122524 classes ($n_\mathcal{P}\approx 1.41 \times 10^{13}$). For 25 men, it is 1 second for $(8,8,8,1)$, which has 2512 classes of pawn arrangements ($n_\mathcal{P}\approx 2.72 \times 10^{9}$) and 1.6 hours for $(6,5,7,7)$, which has 12694182 classes ($n_\mathcal{P}\approx 3.87 \times 10^{14}$).

The upper bound for legal diagrams with 2 to 23 men is increasing with the number of men, up to $5.66\times 10^{35}$ for 23 men. Summing over all these cases, the upper bound is $6.68\times 10^{35}$. This is approximately 10\% better than  respectively  $6.29\times 10^{35}$ and  $7.43\times 10^{35}$ obtained with the method described in \cite{r5}. Preventing the kings to occupy adjacent squares explains this difference: as an argument, consider that $64\times 63 = 4032$ is the total number of ways of placing kings on an empty board, but among them there is only $3612$ legal diagrams, which is approximately 10\% lower.

\section{Conclusion}\label{s4}
\subsection{Further improvements}\label{s41}

The argument given in this paper is not far from optimal. It seems difficult to obtain a significant improvement on the number of legal pawn arrangements: some of the computed classes may be illegal, but they seem to be very few, as none was detected from an examination of 100 randomly chosen classes of 25 men quadruplets. It could also be that inside a given class, some pawn arrangements are illegal. For example, consider the class of $(8,8,8,6)$ in Table \ref{t5}. Only two black pawns have been captured. A simple retrograde analysis reveals that either the white pawn on the c file has taken two black pawns and is now on the e file, or it has taken a pawn on the d file and the white pawn initially on the d file has taken a black pawn on the e file. In any case, any diagram with 3 white pawns on $d2$, $e2$ and $e3$ is illegal. Consequently, the number of pawn arrangements of this class could be bounded by $(6\times 15 -1)\times 15^5\times 6$ instead of $15^6\times 6^2$ given by our method. We believe that this kind of possible improvement has very little effect because few classes are involved and even in those classes the correction is probably low. Similarly, no significant gain would be achieved by computing the list of classes of pawn arrangements for 23 men and less, as one can infer from Table \ref{t4}. 

\begin{table*}[h]
\caption{A class of $(8,8,8,6)$} \label{t5}
\begin{center}
\begin{tabular}{||c c c c c c c c||} 
 \hline
 Col 1 & Col 2 & Col 3 & Col 4 & Col 5 & Col 6 & Col 7 & Col 8 \\ [0.5ex] 
 \hline\hline
 b & b & b & w & w & b & b & b\\ 

 w & w &  &  & w & w & w & w\\ 
 \hline

\end{tabular}
\end{center}
\end{table*}

Regarding piece arrangements, our method takes into account light squared and dark squared bishops for any number of men  and forbids kings to occupy adjacent squares for 23 or less men on the board. We estimate that we would gain approximately 10\% by extending this last rule to 24 men or more, but designing a method achieving that task may prove difficult. As for the bishops, the method we use in subsection \ref{s23} is not optimal, but the loss is only 13\% in the worst case. Of course we also account for a lot of other illegal diagrams, most of which probably with both kings in check. However, we guess that these illegal diagrams are a small minority of the total. For these reasons, we conjecture that the number of legal diagrams in the game of chess without promotions is between $10^{37}$ and $3.5\times 10^{37}$, probably close to $3\times 10^{37}$.

\subsection{Possible extensions}

As explained in Subsection \ref{s41}, it may prove difficult to improve the upper bound we obtained. However, following Tromp (2021b), one could make a ranking of diagrams generated by our work which allows them to be sampled at random, check their legality, and thus arrive at an accurate estimation of the number of legal diagrams without promotion.

Another possible continuation of this work would be to compute an upper bound for the number of legal diagrams allowing pawn promotion. For that purpose, supplementary vertices and edges have to be introduced in the graph described in subsection \ref{s21}. Consider for example vertex $(8,7,8,8)$. Only a black piece is missing on the board. Depending whether a piece or a pawn has taken it, 0, 1 or 2 promotions are possible. This means that this quadruplet has to be partitioned in several vertices, some of them having a possible transition to $(8,8,8,7)$ and to $(9,7,7,8)$ (respectively black and white promotion). However, in this case the majority of arbitrary legal chess diagrams have most likely a lot of promotions and some captures, as it is true for chess positions \cite{r6}. For this reason, it is not obvious to decide whether such a treatment could easily be automated and implemented, in particular regarding storage capacities and computation time. 

Another idea would be to try to adapt this work to study the number of positions, taking into account castling rights, en passant squares and player to move.

\subsection{Verifiability}

Programs and data have been generated using MatLab on a standard desktop computer. Data are available upon request and source code for computing the upper bound for 24 to 32 men is available on github at https://github.com/DanielGourion/ChessDiagrams

\vspace{0.5cm}

\noindent\textbf{ACKNOWLEDGEMENT.} We would like to thank the two anonymous reviewers for their helpful and valuable suggestions and remarks.

\nocite{*}
% if your bibliography is in bibtex format, use those commands:
%\bibliographystyle{ios2-nameyear}  % Style BST file.
%\bibliography{bibliography}        % Bibliography file (usually '*.bib')

% or include bibliography directly:

\end{document}